# Metallic hydrogen with a strong electron-phonon interaction at a pressure of 300-500 GPa

N.N. Degtyarenko, E.A. Mazur [a)] , K.S. Grishakov

National Research Nuclear University MEPhI (Moscow Engineering Physics Institute), Kashirskoe sh., 31, Moscow 115409, Russia

Atomic metallic hydrogen with a lattice with FDDD symmetry is shown to have a stable phase under hydrostatic compression pressure in the range of 350-500 GPa. The resulting structure has a stable spectrum regarding the collapse of the phonons. Ab-unitio simulation method has been used to calculate the structural, electronic, phonon and other characteristics of the normal metallic phase of hydrogen at a pressure of 350-500 GPA.

**1.Introduction.** The theoretical investigations of the metallic hydrogen properties were carried out in works [1-4]. The normal properties of the various crystalline phases of the metallic hydrogen have been calculated. There are prospects to use the atomic hydrogen as a high temperature superconductor as well as a substance with a high density of the stored energy. These prospects are extremely interesting in the issue of producing atomic hydrogen under normal conditions. The normal properties of the I41/amd atomic metallic hydrogen phase at a pressure of 500 GPa were studied in [5]. The superconducting properties of this stable I41/amd phase, predicted in [5], were investigated in [6]. The superconducting transition temperature Tc=217 K of the metallic hydrogen at a pressure of 500 GPa has been obtained in the calculations. The discovery of the metallic hydrogen in the experimental conditions at a pressure of 500 HPa has been reported in [7].

Some phase transformations have been predicted under the compression of the molecular hydrogen on the basis of theoretical calculations. The molecular phases with the symmetries P63/m, C2c, Cmca-12 and Cmca are shown to be stable in the pressure range with the lower pressure values 70, 70÷165, 165÷260, and the higher pressure values of 260 GPa, respectively, in the calculations by the DFT LDA method [8]. The DFT-GGA-PBE method has been used for the calculation of phase diagrams in this work [8] too. It was proved that the phases P63/m, C2c Cmca-12 and Cmca would be stable for the following pressure values: smaller than 110, in the pressure range 110÷245, 245÷370, and when the pressure value is greater than 370 GPA, respectively. These results coincide with calculations [9]. Phase transitions under pressure have been observed experimentally using the changes of both Raman scattering spectra and IR spectra. Direct information was not obtained about the phase structure in these experiments. The pressure value at which reflection spectra changes corresponds to the calculated phase transitions pressure value, for example, for the pressure of P=110 GPa [10]. The range of stability of the I41/amd atomic hydrogen phase (480÷1000 GPa) as well as of the R-3m phase (1000÷1500 GPa) has been computed in [11]. According to [11], the transition of the molecular phase in atomic one should occur near P=500 GPa. From the results of [12] it follows that the metallization of this system occurs at a pressure of about 400 GPa, i.e. there is a range of pressures exceeding 400 GPa where the molecular phase may be metallic.

The question may be asked: what is the pressure P value at which the crystal lattice dynamic stability still is stored when taking the atomic structure as initial with the lowering of the pressure? Such a process corresponds to the movement along the «path 2» in the opposite direction at the phase diagram in the terminology of [7]. In our previous work [5] it is shown that for the phase with I41/amd symmetry the pressure value less 450 GPa is a critical one for the stability of the crystal structure of the metallic hydrogen. In this paper, we have been found the atomic crystal structure phase with the other symmetry. This symmetry marked FDDD is shown to be dynamically stable up to the pressures of the order 350 GPa and in addition has a significantly larger constant of the electron-phonon interaction. This phase is similar with the phase with the I41/amd symmetry.

In accordance with the calculations [1-4], the atomic hydrogen phase may be metastable upon decompression up to the atmospheric values. The possibility of creating metastable atomic hydrogen phase dynamically stable at the lower pressures is a matter for the further research.

**2.Method.** The calculations of structural, electronic and phonon characteristics of the crystal hydrogen structures at a pressure of 500 GPa have been carried out using ab-initio method. The DFT method is used with the plane wave basis as well as the GGA – PBE (Perdew-Burke-Ernzerhof) correlation functional with the pseudopotential that preserves the norm. All the calculations are performed in the spin-polarized approximation with the purpose of adequate comparison of the calculated energy values of different phases. The super cell method is used with the periodic boundary conditions to simulate the crystal. DFPT method with an extended number of unit cells has been used in the calculations of the phonon spectra.

**3.Results. The geometry of the structure.** The rectangular basic cell of the structure with the FDDD symmetry is shown on Fig.1.a. The face-centered orthorhombic primitive cell for this structure (1.b), the reciprocal cell of the primitive cell (fig 1.c) and the isosurfaces of the two e-bands that cross the Fermi level are shown (fig 1.d). There should be seen that the faces of reciprocal cell have a relatively smaller area in the reciprocal space near the points X, Z (Fig.1.c). The path connecting points of high symmetry (G-X-Y-Z-G) is represented by the red lines (fig 1.c):

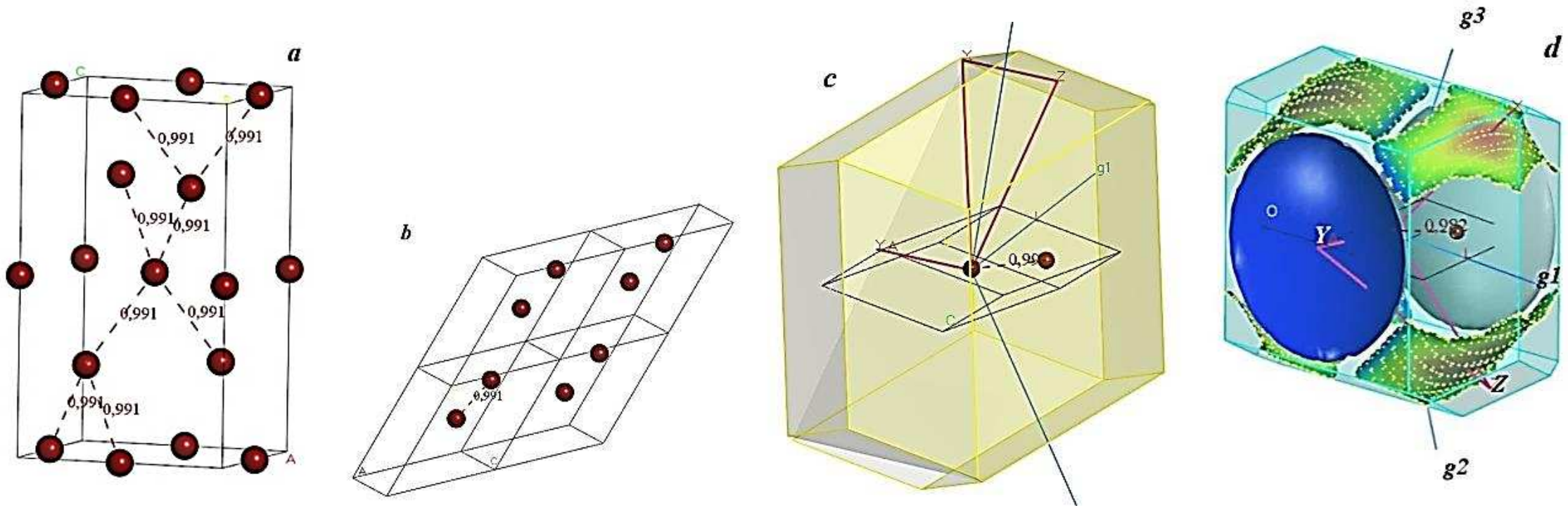

Fig.1. a: Rectangular unit cell of the structure with the FDDD symmetry; b: four primitive orthorhombic cells of the structure shown in Fig.1a; c: inverse cell for the primitive cell, shown in Fig.1b; d: isosurfaces of the two bands crossing the Fermi level.

4. **The parameters of the structures.** Structure with the FDDD symmetry is slightly different from the I41/amd structure studied by us earlier [5]. Rectangular cell (Fig. 1. a) contains 8 atoms, in comparison with the same cell of the I41/amd structure containing 4 atoms. The square face lies at the base of the cell with the I41/amd symmetry. The basic cell structure should be imagined as a rectangular cell structure with the FDDD symmetry with different length of its sides. The difference of the sides of the rectangle base is about 10%, remaining at all pressures. Primitive cells of both structures contain two H atoms. The configuration of H atoms in the first coordination sphere may be considered as the common feature of these two structures. Table.1 presents some of the results of calculations of these structures.The method of calculation and all the parameters have been strictly maintained for these structures. The calculations are performed for the primitive cells. The values of the enthalpies of the two phases without taking into account the phonon contribution are given in table.1.

**Табл.1**

**Structure with FDDD symmetry**

| P [GPa] | *j* | H [eV/atom] | E [eV/atom] | V [A$^3$/atom] | $d_{1min}$ [A] | $d_2$ [A] | $n_{e1}$ | $n_{e2}$ |
|---|---|---|---|---|---|---|---|---|
| 750 | 2 | -8,1318 | -12,6771 | 0,971 | 0,952 | 1,128 | 1,21 | 0,65 |
| 500 | 2 | -9,7705 | -13,3441 | 1,145 | 0,991 | 1, 220 | 1,12 | 0,53 |
| 450 | 2 | -10,1352 | -13,4857 | 1,193 | 1,003 | 1,238 | 1,03 | 0,50 |
| 400 | 2 | -10,5185 | -13,6400 | 1,250 | 1,014 | 1,264 | 0,99 | 0,46 |
| 350 | 2 | -10,9188 | -13,7929 | 1,316 | 1,025 | 1,293 | 0,96 | 0,43 |
| 300 | 2 | -11,3410 | -13,9484 | 1,407 | 1,041 | 1,318 | 0,93 | 0,39 |
| 250 | 2 | -11,7911 | -14,1173 | 1,491 | 1,052 | 1,371 | 0,88 | 0,35 |
| 200 | 2 | -12,2754 | -14,2909 | 1,615 | 1,068 | 1,422 | 0,83 | 0,32 |

**Structure with symmetry I41/AMD**

| P [GPa] | *j* | H [eV/atom] | E [eV/atom] | V [A$^3$/atom] | $d_{1min}$ [A] | $d_2$ [A] | $n_{e1}$ | $n_{e2}$ |
|---|---|---|---|---|---|---|---|---|
| 750 | 2 | -8,1312 | -12,6823 | 0,972 | 0,948 | 1,134 | 1,20 | 0,63 |
| 500 | 2 | -9,7719 | -13,3478 | 1,146 | 0,992 | 1, 206 | 1,08 | 0,62 |
| 450 | 2 | -10,1372 | -13,495 | 1,196 | 0,999 | 1,233 | 1,03 | 0,48 |
| 400 | 2 | -10,5188 | -13,6417 | 1,251 | 1,013 | 1,255 | 0,99 | 0,45 |
| 350 | 2 | -10,9189 | -13,7916 | 1,315 | 1,028 | 1,278 | 0,96 | 0,43 |
| 300 | 2 | -11,3396 | -13,9419 | 1,390 | 1,048 | 1,301 | 0,87 | 0,40 |
| 250 | 2 | -11,7920 | -14,1225 | 1,494 | 1,050 | 1,370 | 0,85 | 0,35 |
| 200 | 2 | -12,2757 | -14,2873 | 1,612 | 1,074 | 1,410 | 0,83 | 0,29 |

Notations: P [GPa] – pressure value; H [eV/atom] is the enthalpy per atom for the primitive cell containing j=2 atoms (the energy of zero-point oscillations is not taken into account); E [eV/atom] is the energy per atom for the primitive cell; V [A3/atom] is the volume per atom of the cell; $d_{1min}$ is the distance to neighboring atoms (first coordination sphere); $d_2$ is the distance to the atoms of the second coordination sphere; $n_{e1}$ is the minimum value of electron density between neighboring atoms; $n_{e2}$ is the minimum value of the electron density on the line to atoms of the second coordination sphere.

The values of the parameters are slightly different for both structures (table 1). Differences of the enthalpy values of both structures lie in the 3rd-4th value after the decimal point. The greatest relative difference corresponds to the magnitude of the distance to the second coordination spheres and to the values of the electron density between the second neighbors, which differ in the second decimal place (of order $10^{-2}$). Although the energy characteristics of these two structures are nearly degenerate, these differences for the second coordination sphere make the structure with the FDDD symmetry dynamically stable at a larger pressure in comparison with the I41/amd phase.

**5. The electronic spectra.** The electronic spectrum of the FDDD phase is represented by two filled bands as it is shown on Fig.2. The states near X and Z points of the reciprocal space are empty. The range of the filled electronic states does not change in the range of pressures 300-500 GPa.

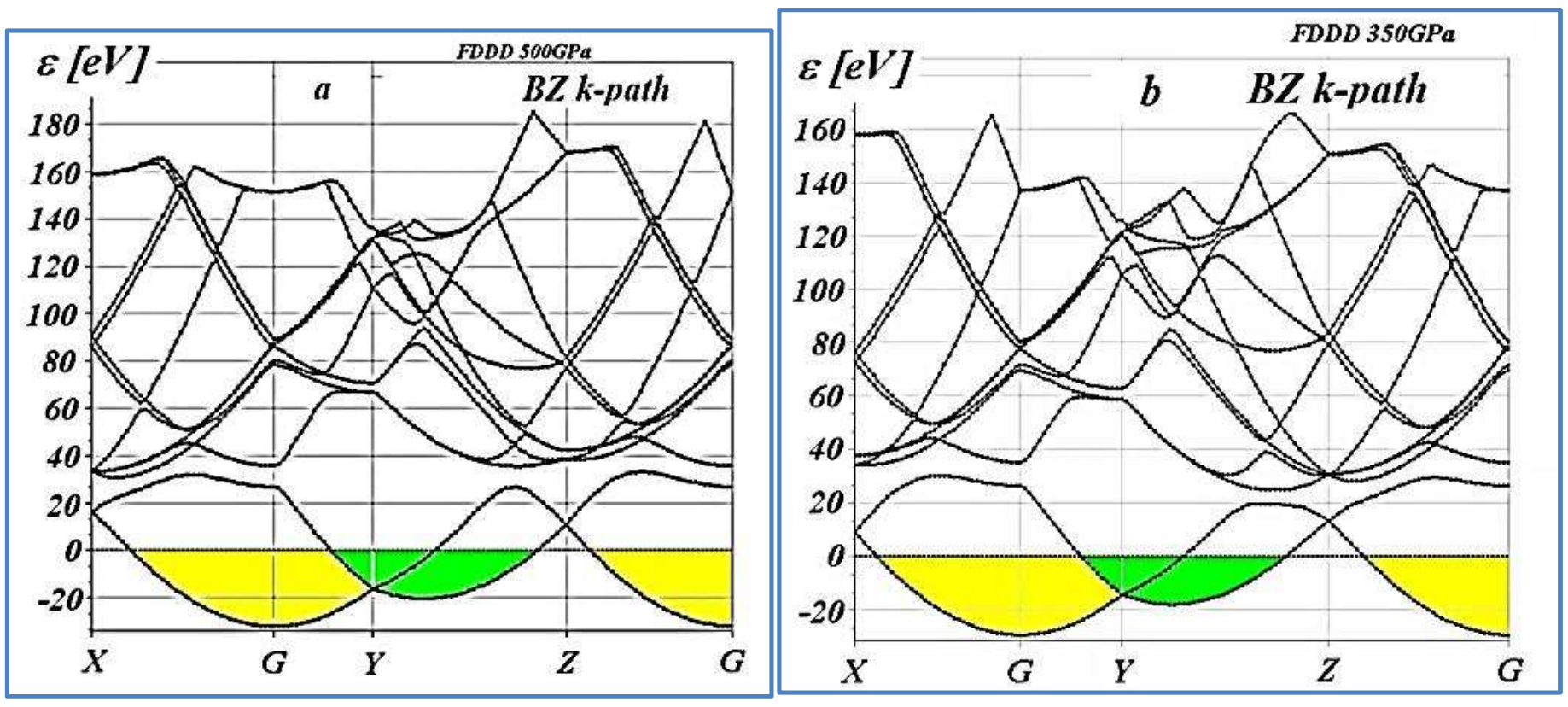

Fig. 2 a, b.The electron band structure for the FDDD phase at pressure of P=500 and 300 GPa, respectively. The calculation for the primitive cell for the structure with j=2 two atoms is presented. Energy ε=0 corresponds to the Fermi level.

The curves of the density of electronic states (DOS) are presented in Fig.3 for the pressures 500 (1), 400 (2) and 300 GPA (3) respectively. There is also shown for the comparison the curve of electronic DOS for the phase with the I41/amd symmetry at a pressure of 500 GPa (1a). The weak change of occupied electronic states is followed from Fig. 3 showing the number of occupied electronic states when the pressure is changing in the range P = 300÷500 GPa. The absolute value of the Fermi energy changes upon compression of the cell with increasing pressure. It has the value 12.27 eV for the pressure of P=300 GPa and increases to 15.42 eV for P=500 GPa (the scale of the graphs does not record these changes).

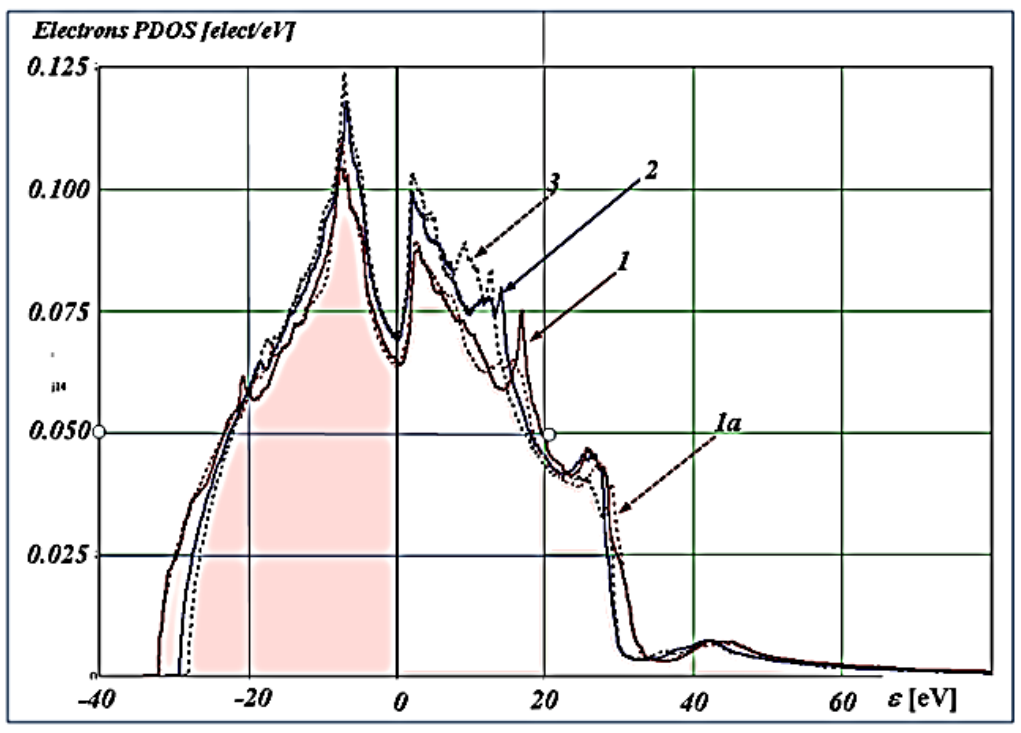

Fig. 3. The density of electronic states DOS for the pressures 500 (1), 400 (2) and 300 ГПа (3) for the primitive cell structure with the FDDD symmetry. Curve (1A) represents DOS at a pressure of 500 HPa for the phase with the I41/amd symmetry.

**6. Phonon spectra.** The changes of the phonon dispersion curves with pressure are more noticeable than the changes of the electron density of states. It is seen from Fig.4 that the imaginary frequencies in the spectrum do not exist for the pressure of P=500 GPa. The imaginary frequencies are also absent for the pressure of 300 GPa along the path (X-G-Y-Z-G). It should be noted that one branche of the oscillations is almost coming to zero frequency on the segment X-Z. In addition, all branches of the calculated spectrum are elevated relative to the zero frequency. The maximum frequency for P=500 GPA corresponds to 370 meV and at the pressure of P=300 GPA corresponds to maximum phonon energy of 300 meV.

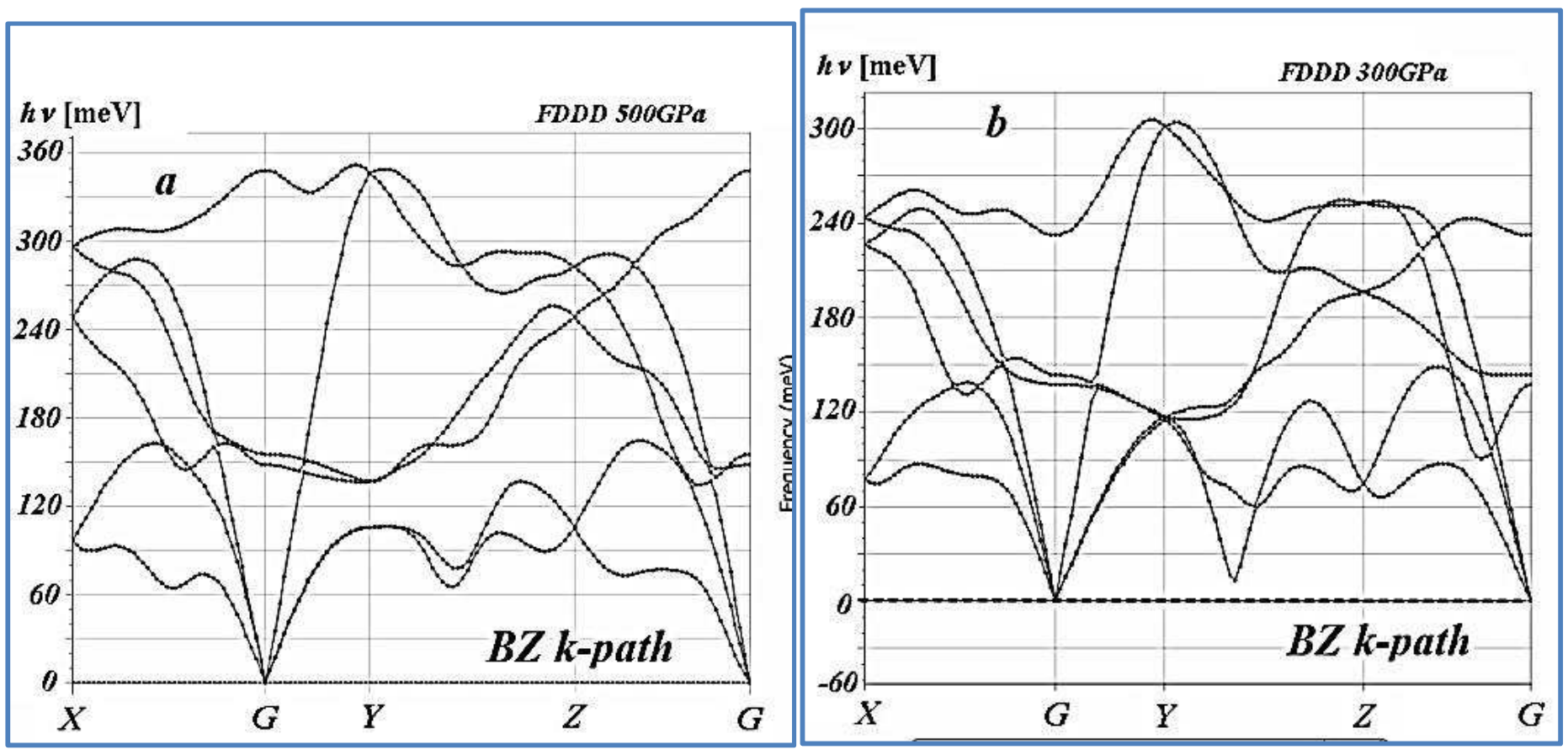

Fig. 4.*a*, *b*. Dispersive phonon dependence (*i=3j=6* is the number of branches of the phonon spectrum) at the pressures of P=500 and 300 GPa, respectively. The calculation of the dynamic matrix by the finite difference method using the super cell of 128 atoms has been carried out for the primitive cell with the number *j=2* of atoms of the FDDD structure.

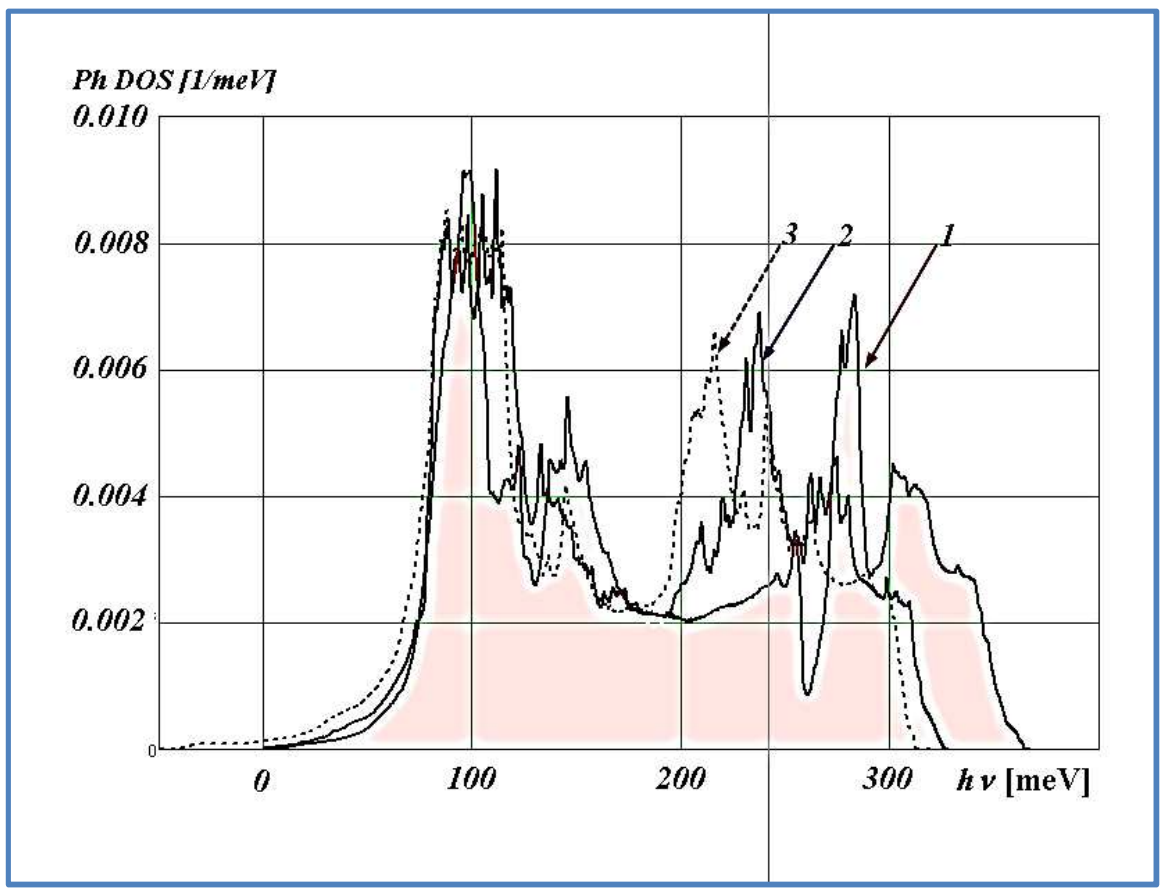

Fig.5. Phonon density of states (PhDOS) for a primitive cell structure with the FDDD symmetry at the different pressures P: 1 – 500, 2 – 350, 3 – 300GPa.

From Fig. 5 is seen that at the pressures of 500 and 350 GPa the imaginary frequencies of the phonon vibrations are absent (the curves come to zero at zero values of the energy). But at 300 GPA there is a "tail" in the region of the "negative" frequencies (the symbol of imaginary frequencies) at zero values of the vibration energy. The qualitative form of the phonon spectrum for the three values of pressure is similar. The spectrum becomes "stiffer", achieving the limiting values of the phonon energy of 370 meV at a pressure of 500 GPa with increasing pressure.

**Zero vibrations.** The dependence of the zero-point energy $E_{ZPE} = (i/2j)\int_0^{\omega_{\max}} h\omega \cdot g(\omega)\cdot d\omega$ from the pressure attributable to a single atom of the unit cell for the structures with FDDD and I41/amd symmetry is shown on fig.6. Here $g_k(\omega)$ is the density of states of each branch of the phonon oscillations with a number of branches from *1* to *i*, normalized to unity. The total density of phonon number fluctuations $g(\omega)$ in all *i* branches of the oscillations also normalized to unity $\int_0^{\omega_{\max}} g(\omega)\cdot d\omega = 1$ is shown in Fig. 5.

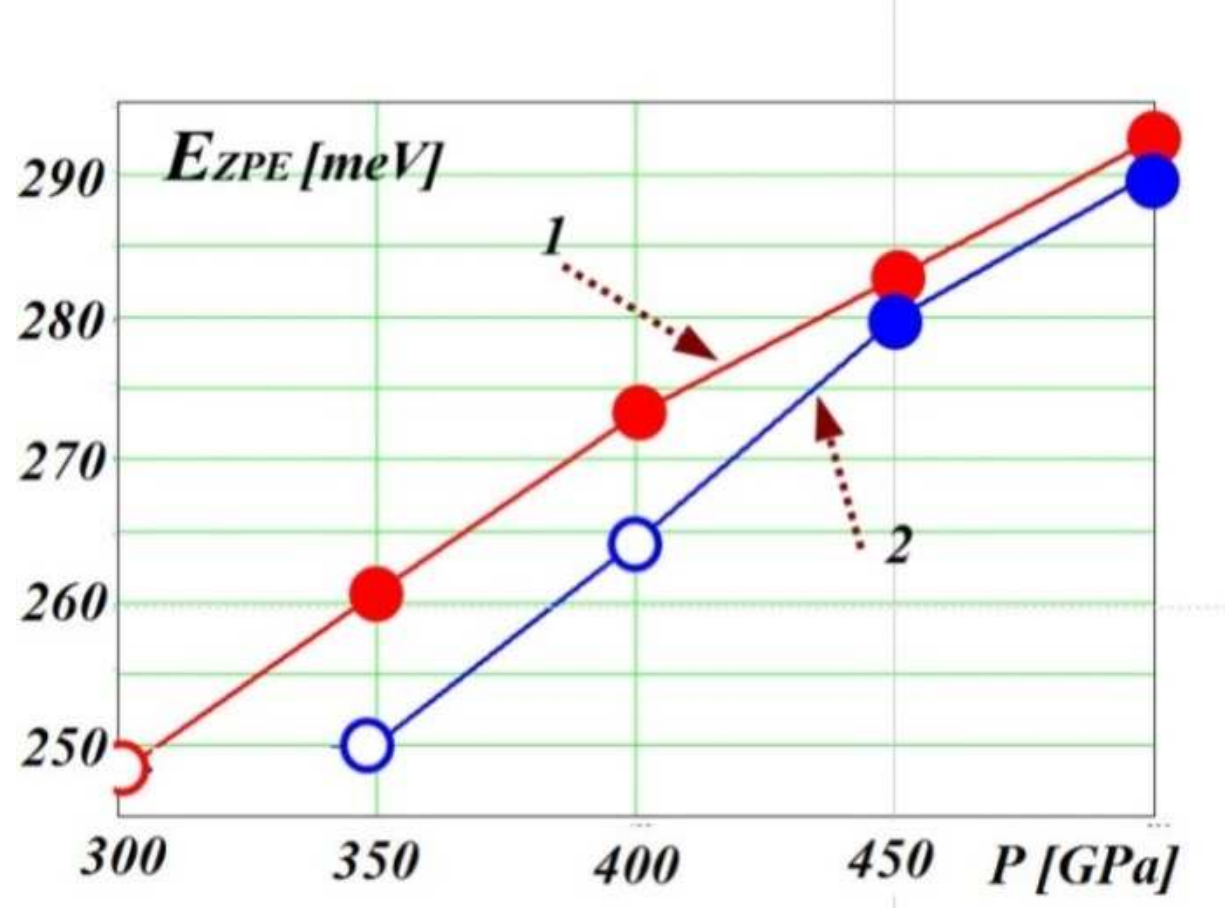

Fig.6. The dependence of the zero-point energy $E_{ZPE}$ per atom vs pressure P for structures with both FDDD (1) and I41/amd (2) symmetry. Empty markers correspond to the presence of imaginary frequencies in the spectrum. The filled markers correspond to the case when the imaginary frequencies are absent, and therefore the structure is dynamically stable.

From the Fig. 6 it is seen that the region of dynamic stability of structures with FDDD symmetry is wider than the region of the dynamic stability of the I41/amd structure. The differences in magnitude of $E_{ZPE}$ for these two structures in the range P=450-500 HPa is of the order of enthalpy difference calculated without taking into account the phonon contribution. Still at the lower pressures the differences in magnitude of $E_{ZPE}$ is growing. As one can see from Fig. 6, the energy of zero-point oscillations per single atom of the structure with FDDD symmetry is greater than the similar energy of the structure with I41/amd symmetry by the amount of 5÷10 meV (50÷100 K).

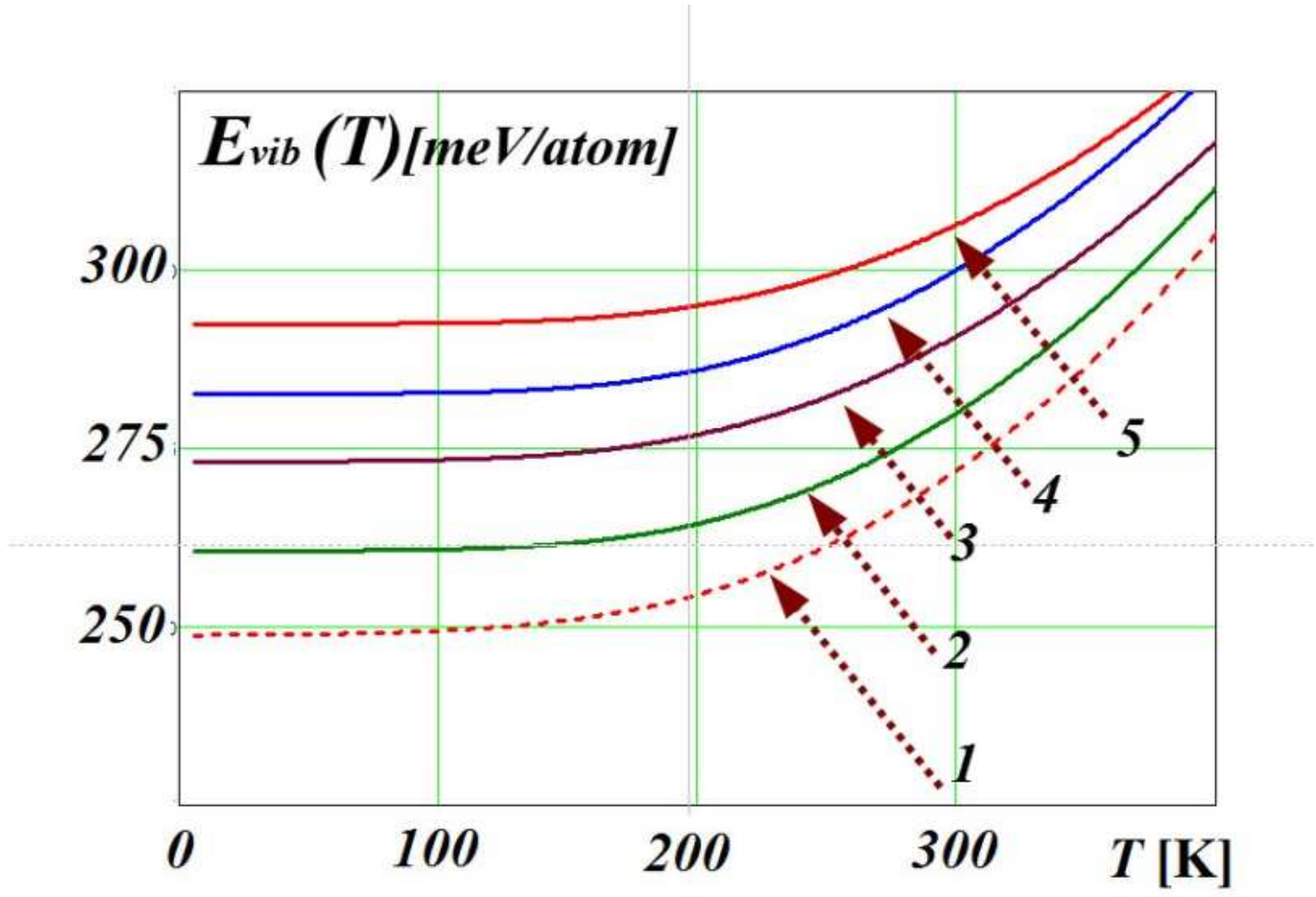

Fig.7. The dependence of the vibrational energy per atom of the structure with the FDDD symmetry vs the temperature for different pressures P: 1 – 300; 2 –350; 3 – 400; 4 –450; 5 – 500 GPa.

The energy of the phonon vibrations per singe H atom is defined as $E_{vib}(T) = E_{ZPE} + (i/j)\int_0^{\omega_{\max}} h\omega / \left(\exp\left(h\omega / T\right) - 1\right) \cdot g(\omega) \cdot d\omega$ . The curves showing this dependency should be seen in Fig.7. It is seen that the effect of the temperature contribution of zero-point oscillations to the background is relatively negligible in the range of the temperatures $T \leq\sim 200K$ . The smallness of the temperature contribution allows us to neglect the additional term $-TS$ in the thermodynamic Gibbs potential $\Omega = U + PV - TS$ . We use in the calculations the enthalpy $H = U + PV$ when considering the problem of the energy stability of any phase under pressure.

**8. Electron-phonon interaction and evaluation of the critical temperature of the superconducting transition.** From Fig.8 it follows that the frequency dependence of the spectral function of electron-phonon interaction $\alpha^2F(\nu)$) is not similar to the frequency dependence of the phonon density of states (PhDOS) at the low frequencies. With the pressure decrease the high-frequency part of the function $\alpha^2F(\nu)$ is shifted to a softer spectrum. The constant of the electron-phonon interaction is defined as $\lambda(\nu) = 2\cdot\int_0^{\nu}(\alpha^2F(x)/x)dx$ . A comparison of the graphs in Fig. 5 and 8 provides insight into the influence of different spectral phonon regions of the structure on a finite value of the constant λ.

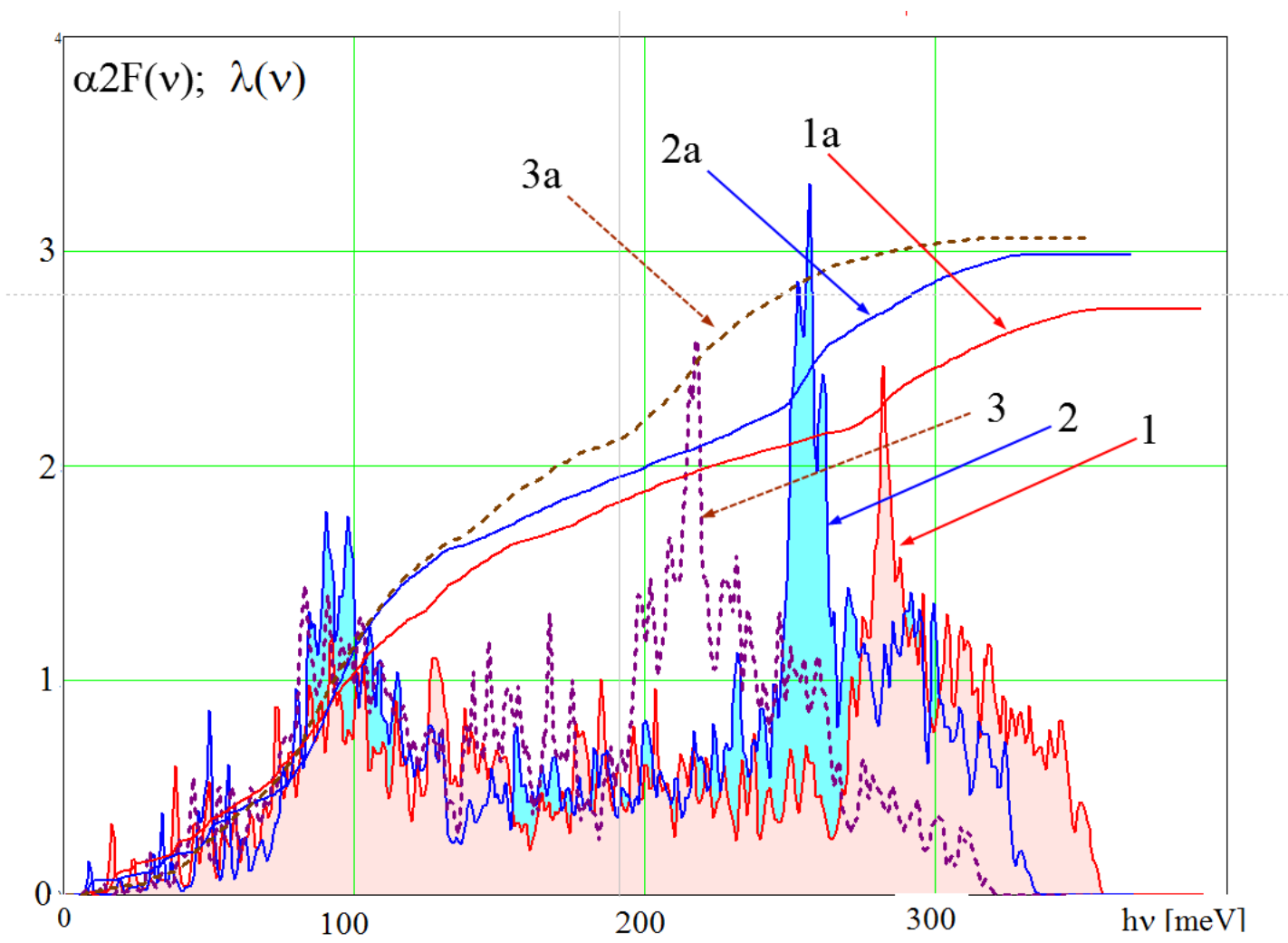

Fig.8. Frequency dependence of the spectral function of electron-phonon interaction $\alpha^2F(\nu)$ and the frequency dependence of the constant of electron-phonon interaction $\lambda(\nu)$ for the structure with FDDD symmetry at the different pressures P: 1, 1a – 500; 2, 2a – 400; 3, 3a – 300 GPa.

The solutions of the Eliashberg equations on the imaginary axis have been carried out using the formalism from [13] for the variable nature of the density of electronic states for the structure with the FDDD symmetry. The following values of λ are obtained: (2.727; 2.983; 3.062), as well as the values of the superconducting gap Δ: (36.7; 39.0; 38.3 meV) for the pressure values P (500; 400 and 300 GPa) respectively. In Fig. 9 the dependence of the critical temperature of the structure with the FDDD symmetry vs. the pressure P is shown.

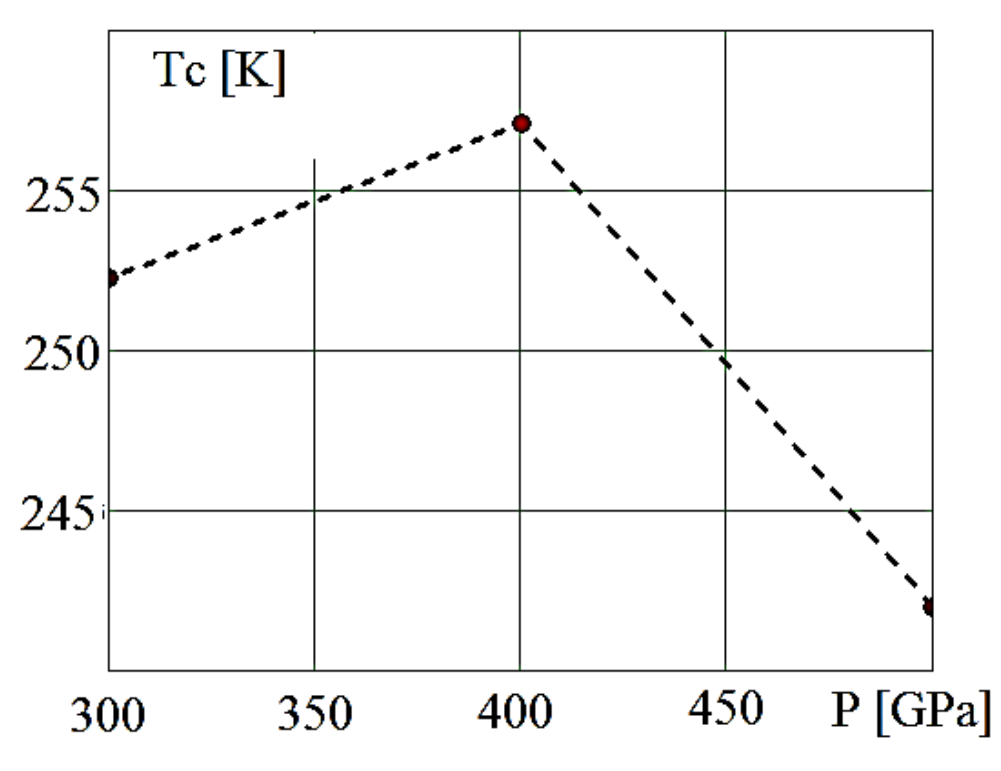

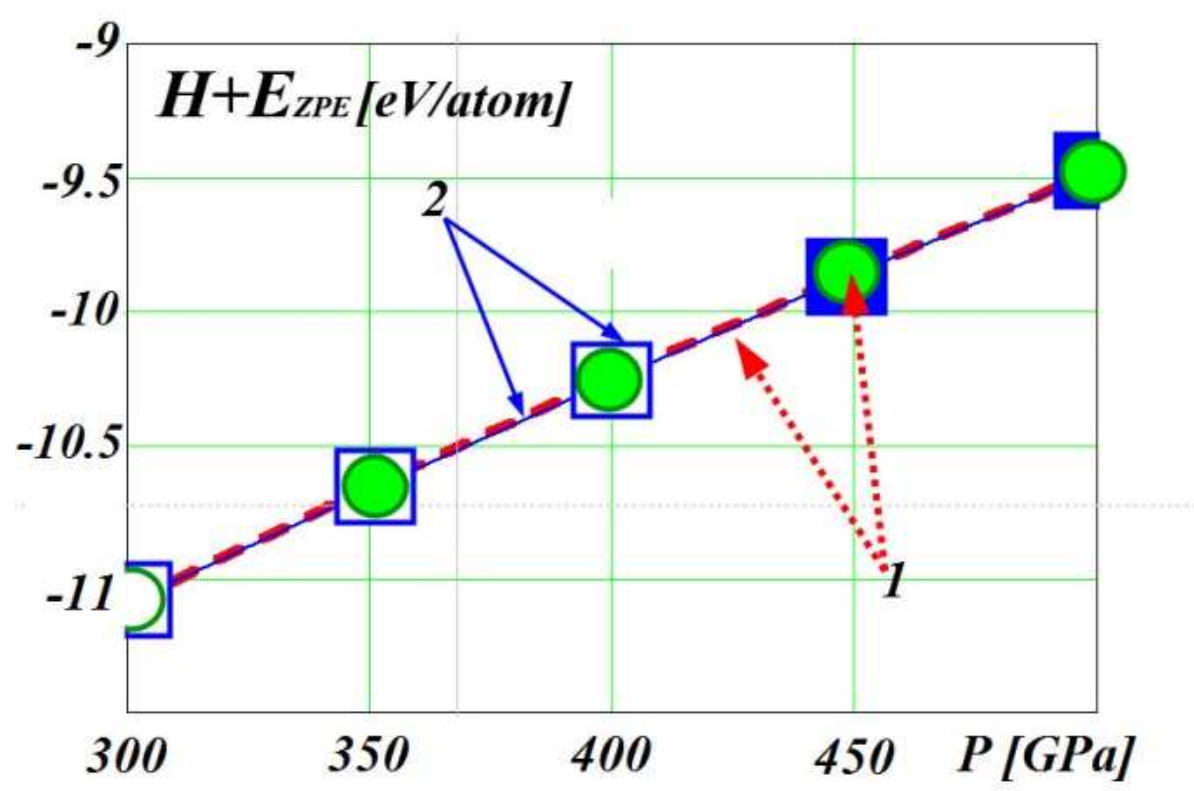

Fig.9. The dependence of the critical temperature of the structure with the FDDD symmetry on the pressure P.

Fig.10. The dependence of the enthalpy with account of the energy of zero vibrations (ZPE) per one atom vs. the pressure P for the primitive cell structures with FDDD symmetry (1– thick dashed line and circular

markers) and I41/amd (2 – thin solid line and square markers). Empty markers correspond to the presence of imaginary frequencies in the spectrum of vibrations.

**9.** Under these calculations, the enthalpy of the FDDD structure excluding the zero-point oscillations is shown to be almost degenerate with the value of the enthalpy of the I41/amd structure (see Tab.1). The addition of the energy of zero-point oscillations, as follows from Fig.10, also does not give a compelling reason to select the most energetically optimal structure. The difference of the values of the enthalpies of the two phases is of the order of 7 meV (80 K) at all pressures. The level of convergence of the iterations and the accuracy of the calculations is of the order of $5 \cdot 10^{-6}$ eV/atom. Such a small difference at temperature of order 200 K will lead to a fluctuational transitions from one metallic hydrogen phase to another one. In [14] the drop of the resistance of a compressed sample of hydrogen at a pressure of 360 GPA is discovered, which may correspond to the percolation effects in the sample with the coexistence of a metallic FDDD phase with the molecular hydrogen phase with the non-metallic character.

A comparison of the zero-point oscillations energy of the crystal structure with the FDDD symmetry with the energy of zero-point oscillations from the calculations [4] of the metallic hydrogen (Fig.4, curve RH2 ) vs the parameter $r_s / a_B$ is shown in Fig.11:

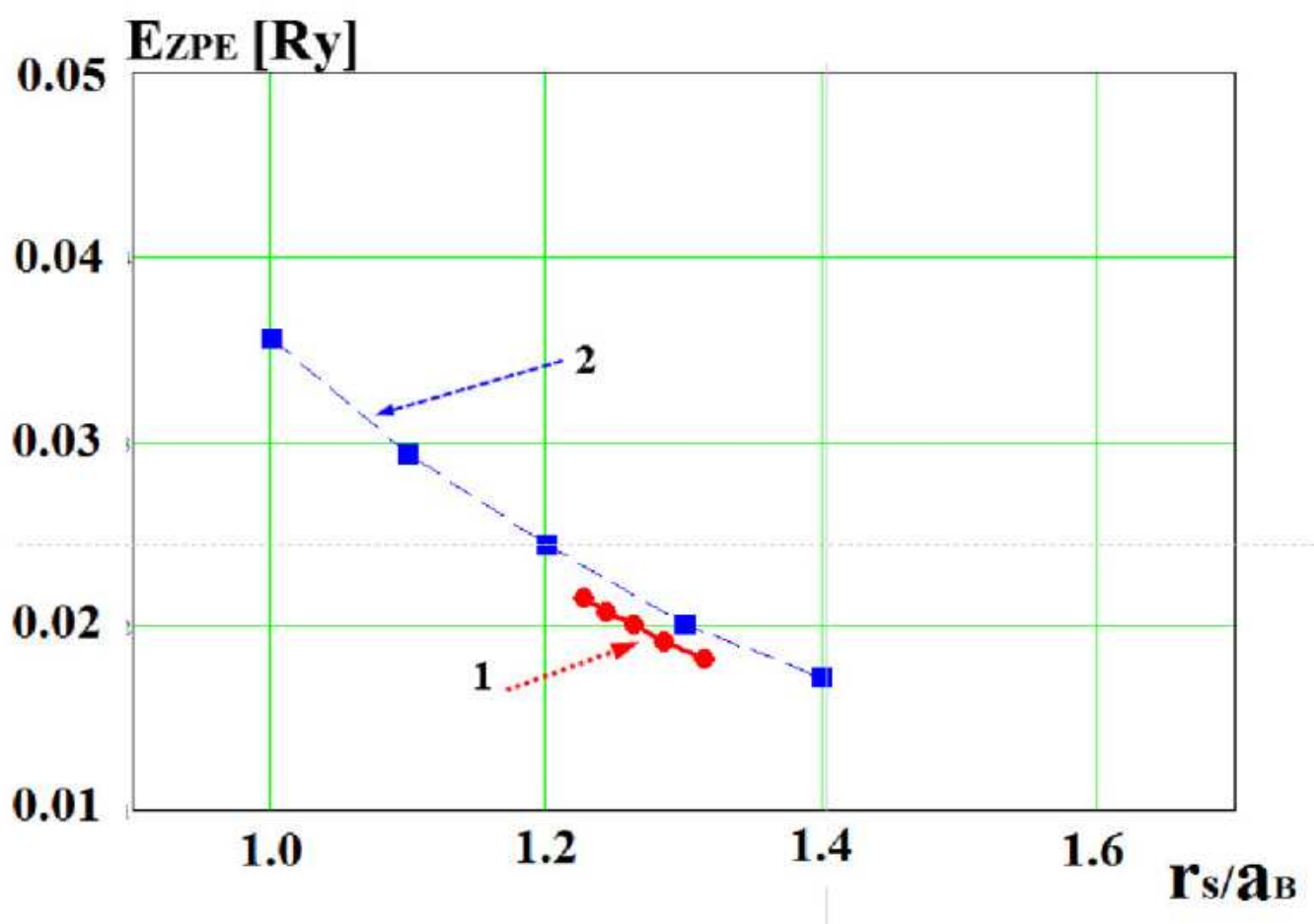

Fig. 11. The energy of the zero phonon oscillations vs. the parameter $r_s / a_B$: (1) – structure with FDDD symmetry, (2) – metallic hydrogen phase from [4]. The 5 points are taken from the curve (Fig.4, curve RH2) of the work [4]. The interpolating curve is carried out through these points.

A comparison of the dependence of both the enthalpy and specific volume per one atom of the structure with FDDD symmetry on the pressure with the results of calculations of metallic hydrogen from [4] is shown in Fig.12:

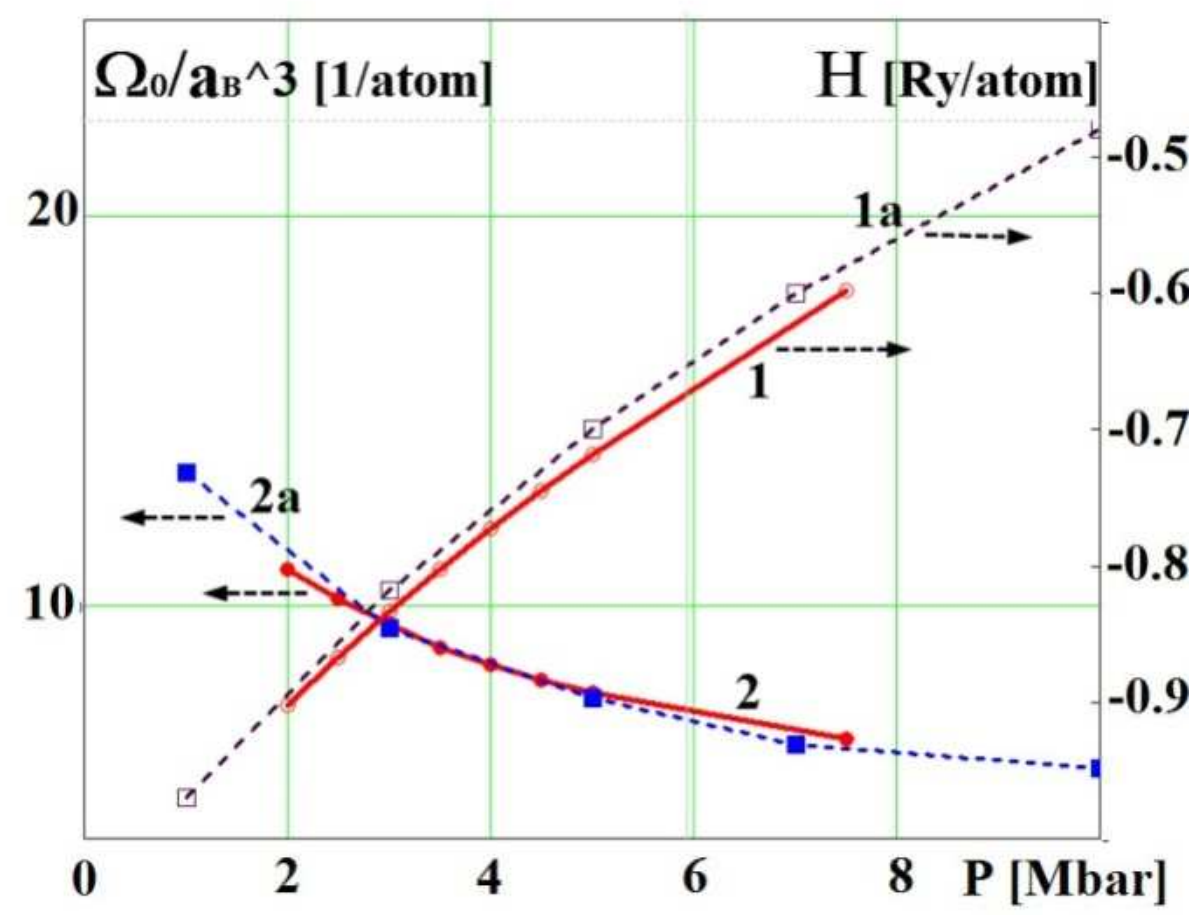

Fig. 12.The comparison of both the dependencies of enthalpy (1) (right vertical axis) and specific volume (2) (left vertical axis) per one atom vs the pressure P for structure with FDDD symmetry with the similar dependencies (1A and 2A) for the metallic hydrogen phase from the work [4]. 5 points are taken from the graphs of the work [4] (Fig.5 – curve 1, Fig.6 - curve 1) through which the interpolating has been conducted marked with the dashed curves.

Good agreement between all the curves describing the equation of state in the form of dependence of specific volume per one atom vs the pressure P, the curve of the energy of zero-point phonon oscillations and the curve of the dependence of the enthalpy per atom vs the pressure P for the FDDD phase with high electron-phonon interaction constant $\lambda \approx 2.7 \div 3$ with the similar results for the monoatomic metallic hydrogen phase [4] points to the universal nature of these dependencies, established in [4], for all superconducting phases of the metallic hydrogen. It is possible to draw a conclusion about a large number of metallic hydrogen phases with close values of the enthalpy coexisting in close proximity to the Tc.

**10. Conclusions.**

1. The structures of the metallic phase of the atomic hydrogen with both FDDD and I41/amd symmetry coexist at temperatures of 150÷250 K due to the small difference of the enthalpy in the pressure interval P=450÷500 GPa with account of the of zero-point oscillations energy.

2. The phase with the FDDD symmetry is shown to be dynamically stable in a wider range of pressures P=350÷500 GPa than the I41/amd phase;

3. $\omega_{max}$ is increasing from the value $\omega_{max}$ =300 meV to the value of $\omega_{max}$ =360 meV during the pressure increase in the interval P=350÷500 GPa.

4. The metallic phase with the FDDD symmetry coexists with a hydrogen molecular metallic phase at the pressures less than 450 GPa.

5. The FDDD phase is characterized by a stronger electron-phonon interaction constant $\lambda \approx 2.7 \div 3$ in comparison with the electron-phonon interaction constant $\lambda \approx 1.8$ in the I41/amd phase [6].

6. Calculation of the critical temperature yields the value of Tc=250 K for the atomic hydrogen with FDDD symmetry of the structure.

7. The universal character of the dependences [4] describing the equation of state of metallic hydrogen, the energy of the zero-point phonon fluctuations and the enthalpy per atom on the pressure for the metallic hydrogen superconducting phases of has been confirmed.

The authors thank Y. Kagan for the discussion of this work.

[a] eugen_mazur@mail.ru